\documentclass[aps,prl,reprint,amsmath,amssymb,superscriptaddress]{revtex4-2}

\usepackage[separate-uncertainty]{siunitx}
\usepackage{graphicx}
\usepackage{braket}
\usepackage[T1]{fontenc}
\usepackage{lmodern}
\usepackage[utf8]{inputenc}
\usepackage{hyperref}

\hypersetup{
	pdfpagemode=UseOutlines,
	bookmarksnumbered=true,
	pdflang=en,
	hidelinks,
	hypertexnames=false,
}

\newcommand{\vect}[1]{\boldsymbol{#1}}
\begin{document}

\title{Spatial imaging of a novel type of molecular ions}

\author{N. Zuber}
\affiliation{5. Physikalisches Institut and Center for Integrated Quantum Science and Technology, Universit\"at Stuttgart, Pfaffenwaldring 57, 70569 Stuttgart, Germany}
\author{V. S. V. Anasuri}
\author{M. Berngruber}
\author{Y.-Q. Zou}
\author{F. Meinert}
\author{R. L\"ow}
\author{T. Pfau}
\affiliation{5. Physikalisches Institut and Center for Integrated Quantum Science and Technology, Universit\"at Stuttgart, Pfaffenwaldring 57, 70569 Stuttgart, Germany}
\date{\today}

\begin{abstract}
Atoms with a highly excited electron, called Rydberg atoms, can form unusual types of molecular bonds~\cite{greeneCreationPolarNonpolar2000,boisseauMacrodimersUltralongRange2002,bendkowskyObservationUltralongrangeRydberg2009a,overstreetObservationElectricfieldinducedCs2009}. The bond differs from the well known ionic and covalent bonds~\cite{bransdenPhysicsAtomsMolecules1986, grayChemicalBondsIntroduction1994a} not only by its binding mechanism, but also by its bond length ranging  up to several micrometres. Here, we observe a new type of molecular ion based on the interaction between the ionic charge and a flipping induced dipole of a Rydberg atom with a bond length of several micrometres. We measure the vibrational spectrum and spatially resolve the bond length and the angular alignment of the molecule using a high-resolution ion microscope~\cite{veitPulsedIonMicroscope2021}. As a consequence of the large bond length, the molecular dynamics is extremely slow. These results pave the way for future studies of spatio-temporal effects in molecular dynamics, e.g., beyond Born-Oppenheimer physics.
\end{abstract}

\maketitle


Ultracold temperatures in atomic and molecular gases have allowed for a new branch of chemistry, where novel weak binding mechanisms between atoms have been observed. To name some examples, these include Feshbach molecules~\cite{kohlerProductionColdMolecules2006}, Efimov states~\cite{greeneUniversalFewbodyPhysics2017a}, few-body collision complexes~\cite{nicholsDetectionLongLivedComplexes2021,mayleScatteringUltracoldMolecules2013}, and ultralong-range Rydberg molecules. The latter include Rydberg ground-state molecules that are formed via  scattering between a highly excited Rydberg electron and a polarisable ground-state atom~\cite{greeneCreationPolarNonpolar2000,bendkowskyObservationUltralongrangeRydberg2009a,shafferUltracoldRydbergMolecules2018,feyUltralongrangeRydbergMolecules2020} and Rydberg macrodimers consisting of two Rydberg atoms bound by a van-der-Waals type of interaction~\cite{boisseauMacrodimersUltralongRange2002,overstreetObservationElectricfieldinducedCs2009}. The long ranging interactions and the corresponding exaggerated bond lengths, exceeding the size of a ground state atom by several orders of magnitude, enable spatially resolved detection by optical microscopy and charged particle optics. These detection schemes allowed also studies of the Rydberg blockade~\cite{lukinDipoleBlockadeQuantum2001a,urbanObservationRydbergBlockade2009a}, Rydberg-Rydberg spatial correlations~\cite{schwarzkopfSpatialCorrelationsRydberg2013,schwarzkopfImagingSpatialCorrelations2011}, many-body states~\cite{schaussObservationSpatiallyOrdered2012,browaeysManybodyPhysicsIndividually2020}, or the examination of the properties of Rydberg macrodimers~\cite{hollerithQuantumGasMicroscopy2019,hollerithMicroscopicElectronicStructure2021b}.

\begin{figure*}[tb!]
\centering
\includegraphics[width=\textwidth]{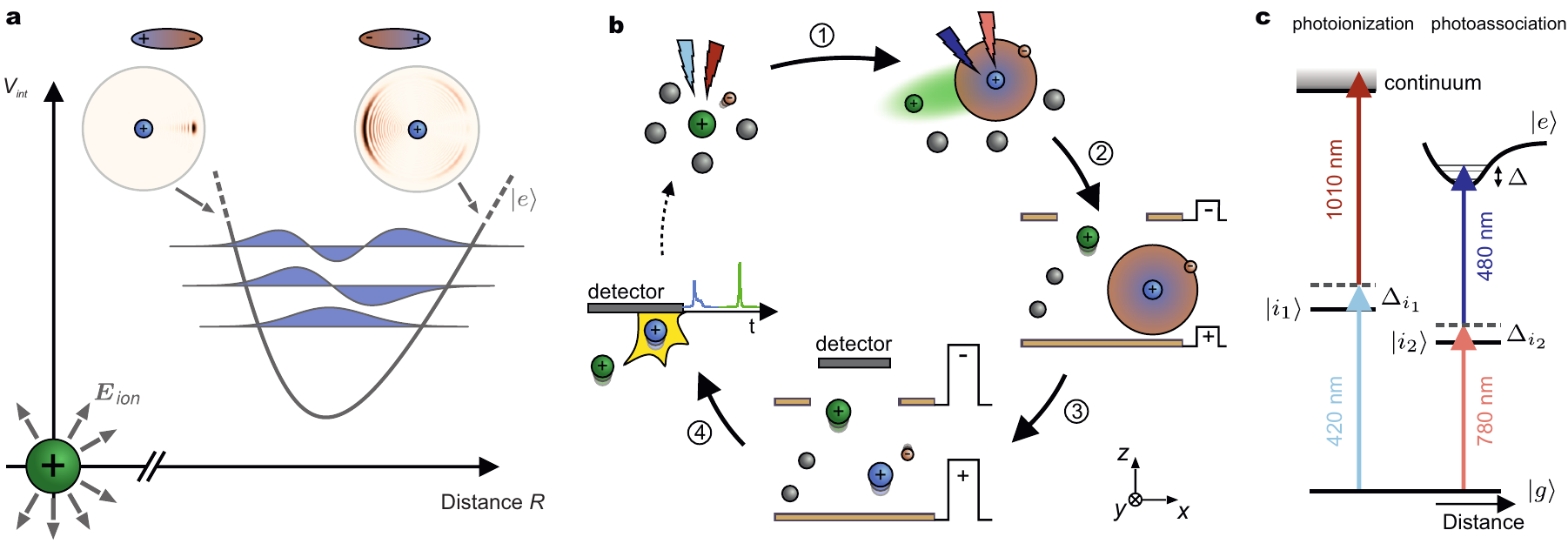}
\caption{\textbf{Binding mechanism of the molecules and experimental sequence.} \textbf{a,} The electric field of the ion $\vect{E}_{\textsf{\textit{ion}}}$ mixes different Rydberg states, leading to a position-dependent polarisation of the Rydberg atom. The electron probability density of the Rydberg electron, visualized above (orange) for two different distances, shows the change from an attractive to a repulsive force. Within the potential curves calculated in Born-Oppenheimer approximation, bound states are formed (blue). \textbf{b,} Measurement cycle showing the creation of an ion, the photoassociation of the molecule, the separation of the Rydberg atom and ion, the extraction, and finally the detection of the ion and the Rydberg core. \textbf{c,} Photoionization and photoassociation process used for molecule creation with a detuning $\Delta_{i_1}$ and $\Delta_{i_2}$ to the intermediate states $\ket{i_1}=\ket{6P_{3/2},F,m_F}$ and $\ket{i_2}=\ket{5P_{3/2},F,m_F}$, respectively.}
\label{fig:concept}%
\end{figure*}

With the advent of experimental techniques enabling the control of cold ions immersed in ultracold neutral atoms~\cite{harterColdAtomIon2014a,kleinbachIonicImpurityBoseEinstein2018,tomzaColdHybridIonatom2019b,ewaldObservationInteractionsTrapped2019b,dieterleInelasticCollisionDynamics2020,schmidtOpticalTrapsSympathetic2020,dieterleTransportSingleCold2021,weckesserObservationFeshbachResonances2021}, the study of weakly-bound neutral molecules can be extended to molecular ions. In this context, a new type of Rydberg molecule has been theoretically proposed, consisting of an ion and a Rydberg atom~\cite{duspayevLongrangeRydbergatomIon2021,deissLongRangeAtomIon2021}. The novel binding mechanism is based on a flipping dipole of the Rydberg atom interacting with the electric field around the ion. As the electronic Rydberg state $\ket{e}$ is polarised by the electric field $\vect{E}_{\textsf{\textit{ion}}}$ of the ion (Fig.~\ref{fig:concept}a), the Rydberg state shifts in energy and forms an interaction potential described here in Born-Oppenheimer approximation. In the case of caesium and rubidium, the quantum defect and the high density of states lead to avoided crossings between the potential curves. Particularly, the $P$-states form potential wells around which the orientation of the dipole can flip, allowing the formation of bound molecular states. Typical bond lengths of this molecular ion reach up to several micrometres and binding depths on the order of gigahertz at these large bond lengths are possible. These properties allow the controlled creation of molecular ions in ultracold gases and the direct observation with ion optics. As for most of these previously described very large and novel molecules, the vibrational dynamics is slowed down dramatically compared to regular molecules, leading to vibrational frequencies in the megahertz range.

Here, we report on the first observation of a molecular ion of that type. We study the vibrational spectrum of the Rydberg-atom-ion molecule, which agrees very well with our theoretical prediction. In addition, we spatially resolve the radial extent and angular alignment of the molecule by an ion microscope~\cite{veitPulsedIonMicroscope2021}. This alignment shows a dependency on the polarisation of the lasers used to photoassociate the molecule. Moreover, we use the spatial resolution of our ion microscope to do time of flight mass spectrometry and show that the excited state is truly a bound molecular state.

The key elements of our experimental procedure to create single molecules and to detect the Rydberg atom and ion individually are shown in Fig.~\ref{fig:concept}b. Each experimental sequence began with an ultracold cloud of unpolarised $^{87}$Rb atoms in the ground state $\ket{g}=\ket{5S_{1/2}, F=2}$ at a temperature of approximately \SI{20}{\micro K}. In one experimental sequence $6{,}080$ measurement cycles were performed. Each measurement cycle starts with the creation of a single cold $\mathrm{Rb}^+$ ion via a two-photon ionization process of a ground state Rb atom. The ionization lasers typically impose an excess energy below $k_B\times\SI{15}{\micro K}$ onto the ion, which is well below the typical binding energy of $\approx k_B\times\SI{100}{mK}$ for vibrational states studied in this paper, where $k_B$ is the Boltzmann constant. Next, a ground state atom can be excited in the vicinity of the ion to the Rydberg state $\ket{e}$ via a two-photon process, but only if the distance between the ion and ground state atom coincides with the bond length  $R$ and if the Rydberg lasers are resonant to the molecular vibrational level(Fig.~\ref{fig:concept}c). To image the spatial structure of the molecule, it is crucial to distinguish the Rydberg atom and ion in our detection scheme. A separate detection of the ion and the Rydberg atom is enabled by an electric separation field pulse, which does not ionize the Rydberg atom but dissociates the molecule and spatially separates the two particles along the optical axis of the ion microscope. After field ionization, this causes a difference in their time of flight through the ion microscope. Consequently, the two particles are distinguishable. The field ionization is initiated by a pulsed electric imaging field that ionizes the Rydberg atom and guides both ion and Rydberg atom core through the ion microscope onto our detector, finishing one measurement cycle. Events comprising the detection of an ion and a Rydberg atom within one measurement cycle are molecular events, whereas the detection of just an ion is referred to as single ion event.

\begin{figure}[t!]
\centering
\includegraphics[width=\columnwidth]{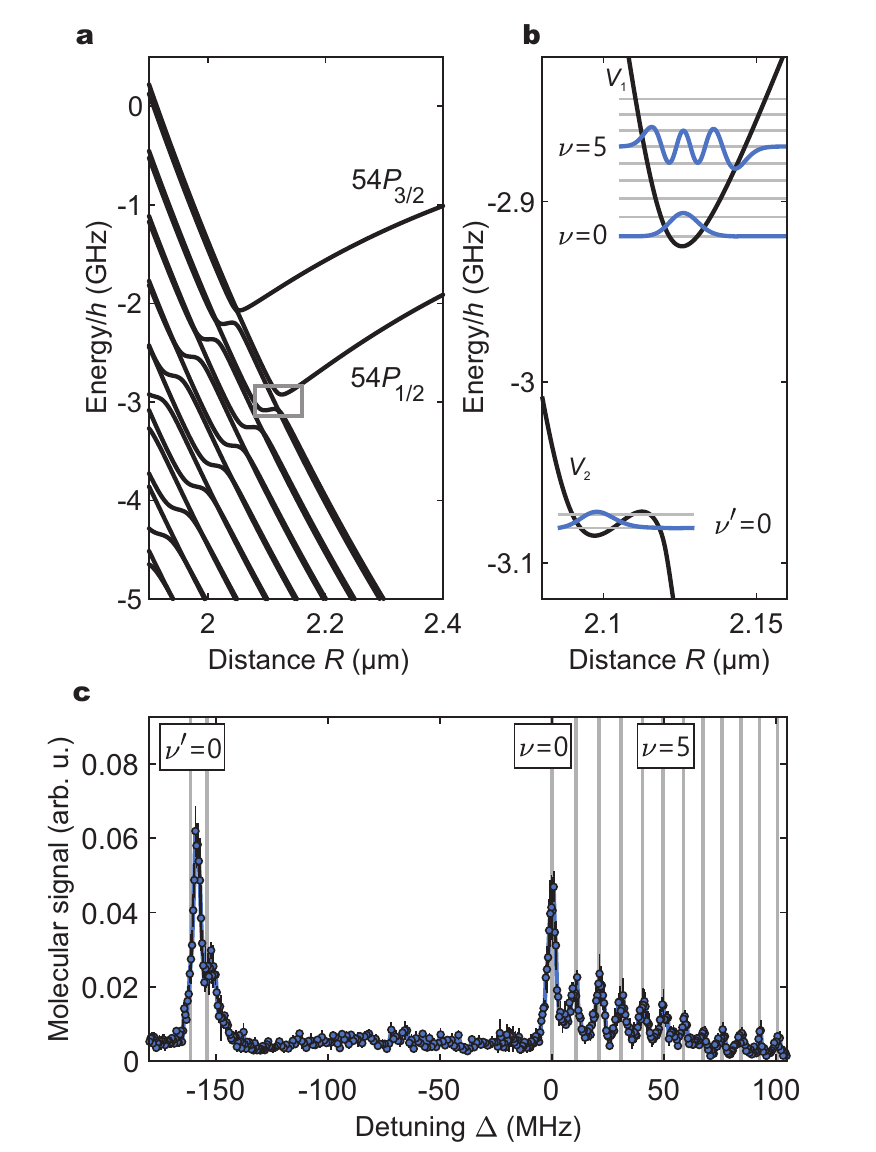}
\caption{\textbf{Calculated potential energy curves with bound vibrational states and measured spectrum.} \textbf{a,} The potential energy curves for the Rydberg-atom-ion molecule corresponding to the $\ket{54P, \left|m_J\right|=1/2}$ states at large distances. \textbf{b,} Magnified view of the region marked with a grey box in \textbf{a} with the two potential curves containing the vibrational bound states. Grey lines indicate the theoretical energies of vibrational states enumerated by $\nu$ and $\nu^\prime$, with vibrational wave functions plotted in blue. \textbf{c,} The normalised molecular signal is shown over the Rydberg laser detuning $\Delta$ and averaged over eight experiment cycles. In the range of \SI{-43}{MHz} to \SI{-122}{MHz} the step size was increased from \SI{0.5}{MHz} to \SI{1}{MHz}. Error bars denote standard error of the mean.}
\label{fig:spectrum}%
\end{figure}

In a first step, we identified the ionic molecule by its vibrational spectrum. An exemplary overview of the theoretically predicted molecular potential curves below the $\ket{54P_{1/2}}$ asymptote is shown in Fig.~\ref{fig:spectrum}a. The methods for calculating the potential are covered in the Methods section and are motivated by recent publications~\cite{duspayevLongrangeRydbergatomIon2021,deissLongRangeAtomIon2021}. The deep outer potential $V_1$ hosts a series of vibrational states (Fig.~\ref{fig:spectrum}b), whereas the shallow inner potential wells ($V_2$ and potentials at smaller $R$) offer a maximum of two bound vibrational states. Due to dipole transition rules, the used two-photon excitation scheme cannot directly excite to a pure $P$-Rydberg state. However, a substantial part of close by $S$- and $D$-states are mixed into the excited Rydberg state (Methods, Fig.~\ref{fig:prefactors69P}).

The shown spectrum was obtained by scanning the detuning $\Delta$ of the \SI{480}{nm} Rydberg excitation laser in \SI{0.5}{MHz} steps (Fig.~\ref{fig:spectrum}c). To reduce the influence of atom number fluctuations in our trap, we normalise the number of molecular events by the amount of detected single ions. Well separated peaks corresponding to the vibrational states are visible in the spectroscopic signal. We attribute the peak marked with $\nu=0$ to the lowest vibrational state of the outer molecular potential well $V_1$. The eleven smaller peaks on the blue-detuned side indicate higher even and odd vibrational levels of the molecule with a maximum vibrational spacing of approximately $\SI{11}{MHz}$. The energies of these vibrational levels are in very good agreement with the theoretically predicted positions (grey line), where the only adjusted parameter is the position of the $\nu=0$ peak. The mean full width at half maximum of the strongest six vibrational states in $V_1$ is $\SI{4.0}{MHz}\pm\SI{0.1}{MHz}$ and exceeds the width of a solely lifetime limited signal, where the measured lifetime of the ground vibrational state is $\SI{11.5}{\micro s}\pm \SI{1.0}{\micro s}$. However, motional transient time broadening effects are estimated to be comparable to the measured linewidth of the vibrational states. Further investigations are necessary to explain the measured lifetime, which is much shorter than the theoretically expected lifetime of the molecule~\cite{duspayevNonadiabaticDecayRydbergatomion2021}. While the cause of this discrepancy is unclear, the lifetime could potentially be reduced by radiofrequency-induced transitions to other unbound Rydberg states. Using a rigid rotor model~\cite{deissLongRangeAtomIon2021}, we estimate the rotational constant to be \SI{26}{Hz} for the states observed in Fig.~\ref{fig:spectrum}c. Consequently, the rotational structure cannot be resolved within the lifetime of the molecule. The two theoretically predicted bound states in the shallow potential well $V_2$ in Fig.~\ref{fig:spectrum}b comply with the two peaks at around $\Delta=\SI{-160}{MHz}$ in the spectrum. Around the avoided crossings at the inner potential curves at $R\approx\SI{1.9}{\micro m}$, excitations of unbound Rydberg atoms are possible and cause a background signal in the spectrum.

\begin{figure}[t!]
\centering
\includegraphics[width=\columnwidth]{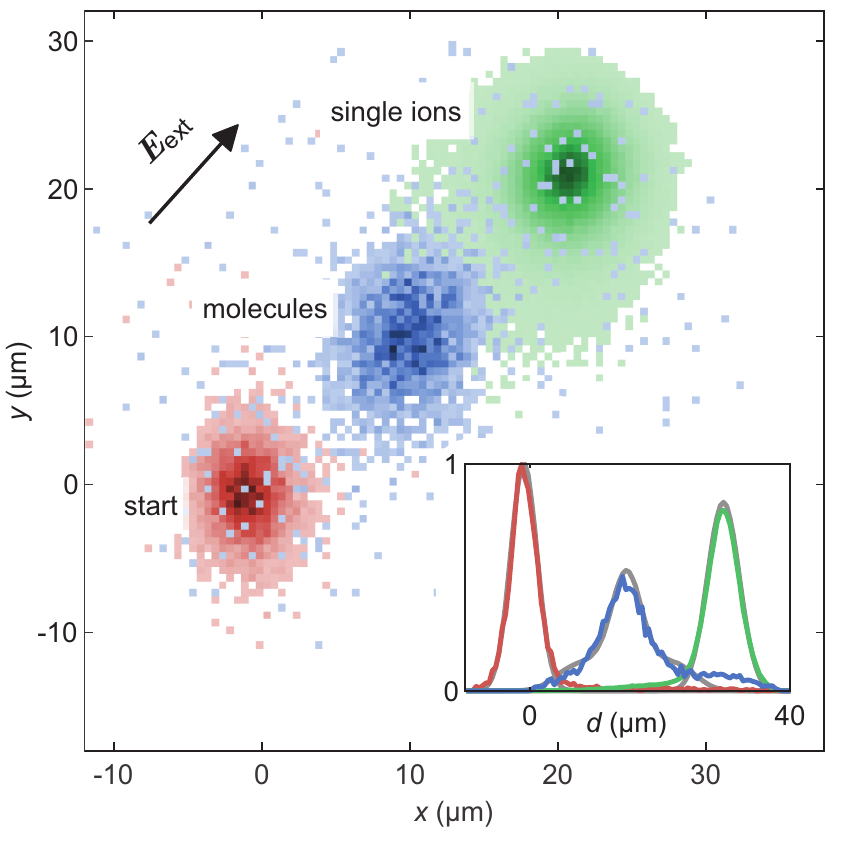}
\caption{\textbf{Mass spectrometry.} Starting from a common region (red), molecules (blue) and single ions (green) are clearly separated when a small external electric field is applied for \SI{12}{\micro s}. The molecular signal contains both, the ion and Rydberg core positions. The origin of the coordinate system is set to the centre of the start region. The separation of single ions and the molecules is a clear indication of bound molecules with double the mass of the single $\mathrm{Rb}^+$ ions. The inset displays the full dataset projected onto the axis parallel to the electric field. Here $d$ denotes the position relative to the start region. Grey lines show the results of classical Monte Carlo trajectory simulations. The curves are individually normalised to the total number of events. The average positions are extracted by Gaussian fits.}
\label{fig:boundstate}%
\end{figure}

To confirm the bound nature of the molecular state, we performed a mass spectrometry via a time of flight for single ions and molecules (Fig.~\ref{fig:boundstate}). The charged particles were accelerated in a constant electric field of $\left|\vect{E}_\textsf{\textit{ext}}\right| \approx \SI{2.3}{mV/cm}$ perpendicular to the optical axis ($z$-axis) of our ion microscope. Consequently, the single and molecular ions travelled different distances in the object plane during an evolution time $t_{e}$ between the photoassociation and the separation pulse due to their different charge to mass ratio. In Fig.~\ref{fig:boundstate}, the initial position for $t_e=\SI{0}{\micro s}$ is depicted in red. For an evolution time of $t_e=\SI{12}{\micro s}$ the $\mathrm{Rb}^+$ ions (green) travelled on average a distance of $\SI{30.6}{\micro m} \pm \SI{0.1}{\micro m}$, whereas the molecules (blue) moved $\SI{15.5}{\micro m} \pm \SI{0.1}{\micro m}$. Classical Monte Carlo trajectory simulations shown in the inset of Fig.~\ref{fig:boundstate} (grey) agree very well with the position for the molecules and the single ions. This result confirms the anticipated bond between the ion and the Rydberg atom. However, not all molecules remained in a bound state during the evolution time as the detected molecules show broad wings in the spatial profile (Fig.~\ref{fig:boundstate}, Inset). To our understanding, this broadening results from dissociated molecules, for which the single ion had accelerated further in the electric field after the dissociation. By post-processing the data, we could confirm that the broadening originated from events with a Rydberg-atom-ion distance of more than \SI{5}{\micro m}, which is larger than the bond length. The dissociation process is included in the simulation by assuming an exponential time dependence with a time scale of $\SI{11.5}{\micro s}$ corresponding to the independently measured lifetime of the molecule. Our theoretical model reproduces the main features in the measured signal well. However, further investigation is necessary to explain the dissociation process.

Our ion microscope~\cite{veitPulsedIonMicroscope2021} provides the capability to image single particles not only with a resolution of \SI{200}{nm}, but allows us to distinguish between ions and Rydberg atoms to create in situ images of the free floating molecular ion. These images complement the presented spectroscopic data of the molecule. For the precise measurement of the bond length and molecular alignment, the Rydberg excitation lasers were tuned to excite the vibrational ground state in the potential below the $\ket{69 P_{1/2}}$ asymptote (see Methods). The interaction strength and binding energy are therefore decreased in comparison to the previously studied molecular state and the dissociation process induced by the separation pulse takes place on a faster time scale.

\begin{figure}[t!]
\centering
\includegraphics[width=\columnwidth]{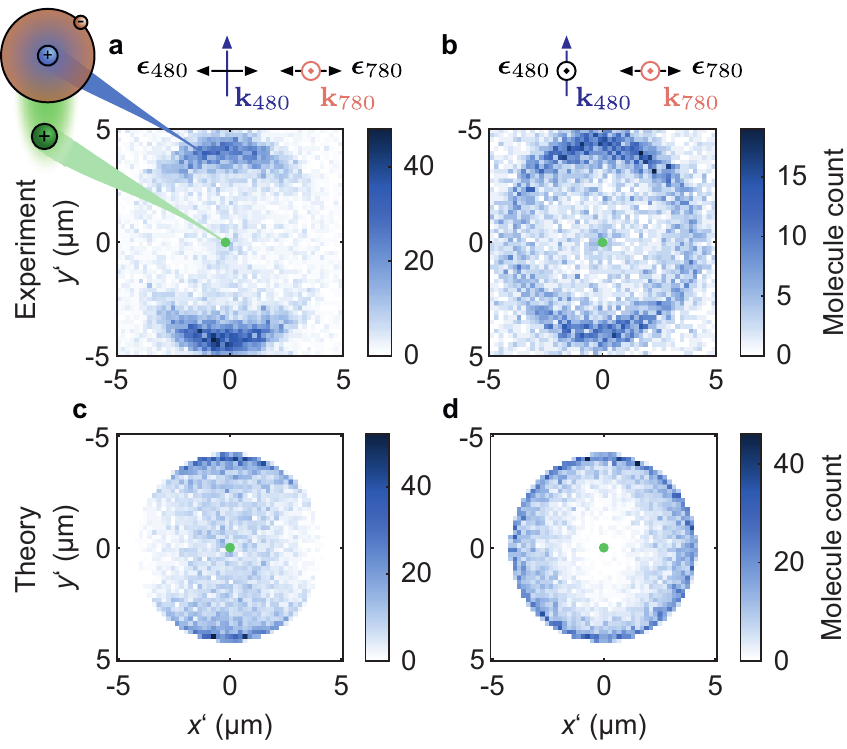}
\caption{\textbf{In situ measurements of the molecule.} \textbf{a} and \textbf{b}, For two configurations of the laser polarisation $\vect{\epsilon}_{480}$ in situ measurements of the molecule are shown. Here, $x^\prime$ and $y^\prime$ represent the relative position of the Rydberg atom with respect to the ion. \textbf{c} and \textbf{d}, The bottom row displays theoretical predictions, where the position of the Rydberg atom is randomly drawn from an angular dependent probability distribution according to relative two-photon excitation probabilities.}
\label{fig:insituimage}%
\end{figure}

The experimental results in panels a and b of Fig.~\ref{fig:insituimage} show about $8{,}000$ and $12{,}000$ molecules, which were imaged \SI{700}{ns} after the photoassociation. For each detected molecule the relative distance and orientation of the Rydberg core and the ion were extracted from the measurement data such that the ion is located at the origin. The most probable radial distance between the ion and the Rydberg core in Fig.~\ref{fig:insituimage}a and b is \SI{4.3}{\micro m} and differs by less than 5\% from the theoretical predicted bond length of around \SI{4.1}{\micro m}.

The dependence of the molecular excitation probability with respect to the laser polarisation direction $\vect{\epsilon}_{480}$, defined in the laboratory frame, is clearly visible. The finite waist of our \SI{480}{nm} Rydberg excitation laser and anisotropies in the atomic cloud density contribute to slight angular asymmetries in the detected molecular signal compared to theoretical predictions (Fig.~\ref{fig:insituimage}c, d). This can be seen for the case shown on the right, where molecules oriented along the $y^\prime$-axis are slightly more probable compared to the $x^\prime$-axis. Here the prime denotes the relative coordinates between the two constituents. When the Rydberg lasers are tuned away from resonance to the vibrational states, the spatial structure visible in Fig.~\ref{fig:insituimage}a and b vanishes.

Theoretical predictions (Fig.~\ref{fig:insituimage}c, d) were obtained by taking into account the angular dependent excitation probability and the radial vibrational wave function of the molecule to randomly sample its orientation. The angular orientation of the molecular ions depends on the two-photon excitation probability per unit time~\cite{boninTwophotonElectricdipoleSelection1984}, which is proportional to $\sum_{m_J,m_F}\left| \sum_{i_2} \Delta_{i_2}^{-1} \braket{e, m_J | \vect{\varepsilon}_{480} \cdot \vect{\hat{d}} |i_2} \braket{i_2 | \vect{\varepsilon}_{780} \cdot \vect{\hat{d}} |g, m_F} \right| ^2 $ (see Methods). Here $\vect{\hat{d}}$ is the dipole operator, $\vect{\varepsilon}_\lambda$ the polarisation vector of the respective Rydberg excitation laser in the molecular frame and $\Delta_{i_2}$ is the laser detuning to the intermediate hyperfine states $\ket{i_2}$. The probabilities are summed up over all the hyperfine states $m_F$ of the ground $\ket{g,m_F}$ state and the fine structure states $\ket{e,m_J}$ of the excited state with magnetic quantum number $m_J$. Broadening effects, like the dispersion of the radial wave packet during the separation pulse or residual effects of the dissociation process, are not included in the theory calculations. However, these influences can be further suppressed in future experiments by changing to heteronuclear molecules, as the different mass of the particles would separate the two particles during the imaging process in the ion microscope, rendering the separation pulse unnecessary.

To conclude, we observed a new type of molecular bond between an ion and a Rydberg atom with our ion microscope. The measured vibrational structure of the novel ionic molecule agrees very well with theoretical predictions and should be observable in other atomic species which exhibit similar quantum defects. The large spatial structure reduces the vibrational and rotational time scales of the molecule. These exotic properties in combination with our high resolution ion detection scheme allow in situ observation of these molecules and possibly enable the direct spatially resolved observation of wave packet dynamics within the molecular potentials. Other new research directions include the observation of trimers or even polymers with multiple correlated bound Rydberg atoms.

\section{Acknowledgements}
We thank T. Schmid and C. Veit for their work and support in building the ion microscope. Further, we thank O.A. Herrera-Sancho for his support during the calibration of the microscope and S. Weber for his help with the pairinteraction code. We are indebted to the mechanics and electronics workshop for manufacturing the ion microscope and the high voltage power supplies. This work was supported by the Deutsche Forschungsgemeinschaft (DFG) [Project No. Pf 381/17-1 and No. Pf 381/17-2] as part of the SPP 1929 "Giant Interactions in Rydberg Systems (GiRyd)" and has received funding from the Bundesministerium für Bildung und Forschung (BMBF) - project "Theory-Blind Quantum Control - TheBlinQC" (grant no. 13N14847). The project received funding from the
European Research Council (ERC) under the European Union’s Horizon 2020 research and innovation programme (Grant agreement No. 101019739 - LongRangeFermi). F. M. acknowledges support from the Carl-Zeiss foundation via IQST and is indebted to the Baden-Württemberg-Stiftung for the financial support by the Eliteprogramm for Postdocs.

%
\bibliographystyle{apsrev4-2}

\begin{thebibliography}{37}%
\makeatletter
\providecommand \@ifxundefined [1]{%
 \@ifx{#1\undefined}
}%
\providecommand \@ifnum [1]{%
 \ifnum #1\expandafter \@firstoftwo
 \else \expandafter \@secondoftwo
 \fi
}%
\providecommand \@ifx [1]{%
 \ifx #1\expandafter \@firstoftwo
 \else \expandafter \@secondoftwo
 \fi
}%
\providecommand \natexlab [1]{#1}%
\providecommand \enquote  [1]{``#1''}%
\providecommand \bibnamefont  [1]{#1}%
\providecommand \bibfnamefont [1]{#1}%
\providecommand \citenamefont [1]{#1}%
\providecommand \href@noop [0]{\@secondoftwo}%
\providecommand \href [0]{\begingroup \@sanitize@url \@href}%
\providecommand \@href[1]{\@@startlink{#1}\@@href}%
\providecommand \@@href[1]{\endgroup#1\@@endlink}%
\providecommand \@sanitize@url [0]{\catcode `\\12\catcode `\$12\catcode
  `\&12\catcode `\#12\catcode `\^12\catcode `\_12\catcode `\%12\relax}%
\providecommand \@@startlink[1]{}%
\providecommand \@@endlink[0]{}%
\providecommand \url  [0]{\begingroup\@sanitize@url \@url }%
\providecommand \@url [1]{\endgroup\@href {#1}{\urlprefix }}%
\providecommand \urlprefix  [0]{URL }%
\providecommand \Eprint [0]{\href }%
\providecommand \doibase [0]{https://doi.org/}%
\providecommand \selectlanguage [0]{\@gobble}%
\providecommand \bibinfo  [0]{\@secondoftwo}%
\providecommand \bibfield  [0]{\@secondoftwo}%
\providecommand \translation [1]{[#1]}%
\providecommand \BibitemOpen [0]{}%
\providecommand \bibitemStop [0]{}%
\providecommand \bibitemNoStop [0]{.\EOS\space}%
\providecommand \EOS [0]{\spacefactor3000\relax}%
\providecommand \BibitemShut  [1]{\csname bibitem#1\endcsname}%
\let\auto@bib@innerbib\@empty
\bibitem [{\citenamefont {Greene}\ \emph {et~al.}(2000)\citenamefont {Greene},
  \citenamefont {Dickinson},\ and\ \citenamefont
  {Sadeghpour}}]{greeneCreationPolarNonpolar2000}%
  \BibitemOpen
  \bibfield  {author} {\bibinfo {author} {\bibfnamefont {C.~H.}\ \bibnamefont
  {Greene}}, \bibinfo {author} {\bibfnamefont {A.~S.}\ \bibnamefont
  {Dickinson}},\ and\ \bibinfo {author} {\bibfnamefont {H.~R.}\ \bibnamefont
  {Sadeghpour}},\ }\bibfield  {title} {\bibinfo {title} {Creation of {{Polar}}
  and {{Nonpolar Ultra}}-{{Long}}-{{Range Rydberg Molecules}}},\ }\href
  {https://doi.org/10.1103/PhysRevLett.85.2458} {\bibfield  {journal} {\bibinfo
   {journal} {Phys. Rev. Lett.}\ }\textbf {\bibinfo {volume} {85}},\ \bibinfo
  {pages} {2458} (\bibinfo {year} {2000})}\BibitemShut {NoStop}%
\bibitem [{\citenamefont {Boisseau}\ \emph {et~al.}(2002)\citenamefont
  {Boisseau}, \citenamefont {Simbotin},\ and\ \citenamefont
  {C{\^o}t{\'e}}}]{boisseauMacrodimersUltralongRange2002}%
  \BibitemOpen
  \bibfield  {author} {\bibinfo {author} {\bibfnamefont {C.}~\bibnamefont
  {Boisseau}}, \bibinfo {author} {\bibfnamefont {I.}~\bibnamefont {Simbotin}},\
  and\ \bibinfo {author} {\bibfnamefont {R.}~\bibnamefont {C{\^o}t{\'e}}},\
  }\bibfield  {title} {\bibinfo {title} {Macrodimers: Ultralong {{Range Rydberg
  Molecules}}},\ }\href {https://doi.org/10.1103/PhysRevLett.88.133004}
  {\bibfield  {journal} {\bibinfo  {journal} {Physical Review Letters}\
  }\textbf {\bibinfo {volume} {88}},\ \bibinfo {pages} {133004} (\bibinfo
  {year} {2002})}\BibitemShut {NoStop}%
\bibitem [{\citenamefont {Bendkowsky}\ \emph {et~al.}(2009)\citenamefont
  {Bendkowsky}, \citenamefont {Butscher}, \citenamefont {Nipper}, \citenamefont
  {Shaffer}, \citenamefont {L{\"o}w},\ and\ \citenamefont
  {Pfau}}]{bendkowskyObservationUltralongrangeRydberg2009a}%
  \BibitemOpen
  \bibfield  {author} {\bibinfo {author} {\bibfnamefont {V.}~\bibnamefont
  {Bendkowsky}}, \bibinfo {author} {\bibfnamefont {B.}~\bibnamefont
  {Butscher}}, \bibinfo {author} {\bibfnamefont {J.}~\bibnamefont {Nipper}},
  \bibinfo {author} {\bibfnamefont {J.~P.}\ \bibnamefont {Shaffer}}, \bibinfo
  {author} {\bibfnamefont {R.}~\bibnamefont {L{\"o}w}},\ and\ \bibinfo {author}
  {\bibfnamefont {T.}~\bibnamefont {Pfau}},\ }\bibfield  {title} {\bibinfo
  {title} {Observation of ultralong-range {{Rydberg}} molecules},\ }\href
  {https://doi.org/10.1038/nature07945} {\bibfield  {journal} {\bibinfo
  {journal} {Nature}\ }\textbf {\bibinfo {volume} {458}},\ \bibinfo {pages}
  {1005} (\bibinfo {year} {2009})}\BibitemShut {NoStop}%
\bibitem [{\citenamefont {Overstreet}\ \emph {et~al.}(2009)\citenamefont
  {Overstreet}, \citenamefont {Schwettmann}, \citenamefont {Tallant},
  \citenamefont {Booth},\ and\ \citenamefont
  {Shaffer}}]{overstreetObservationElectricfieldinducedCs2009}%
  \BibitemOpen
  \bibfield  {author} {\bibinfo {author} {\bibfnamefont {K.~R.}\ \bibnamefont
  {Overstreet}}, \bibinfo {author} {\bibfnamefont {A.}~\bibnamefont
  {Schwettmann}}, \bibinfo {author} {\bibfnamefont {J.}~\bibnamefont
  {Tallant}}, \bibinfo {author} {\bibfnamefont {D.}~\bibnamefont {Booth}},\
  and\ \bibinfo {author} {\bibfnamefont {J.~P.}\ \bibnamefont {Shaffer}},\
  }\bibfield  {title} {\bibinfo {title} {Observation of electric-field-induced
  {{Cs Rydberg}} atom macrodimers},\ }\href {https://doi.org/10.1038/nphys1307}
  {\bibfield  {journal} {\bibinfo  {journal} {Nature Physics}\ }\textbf
  {\bibinfo {volume} {5}},\ \bibinfo {pages} {581} (\bibinfo {year}
  {2009})}\BibitemShut {NoStop}%
\bibitem [{\citenamefont {Bransden}\ and\ \citenamefont
  {Joachain}(1986)}]{bransdenPhysicsAtomsMolecules1986}%
  \BibitemOpen
  \bibfield  {author} {\bibinfo {author} {\bibfnamefont {B.~H.}\ \bibnamefont
  {Bransden}}\ and\ \bibinfo {author} {\bibfnamefont {C.~J.}\ \bibnamefont
  {Joachain}},\ }\href@noop {} {\emph {\bibinfo {title} {Physics of Atoms and
  Molecules}}}\ (\bibinfo  {publisher} {{Longman}},\ \bibinfo {address}
  {{London ; New York}},\ \bibinfo {year} {1986})\BibitemShut {NoStop}%
\bibitem [{\citenamefont {Gray}(1994)}]{grayChemicalBondsIntroduction1994a}%
  \BibitemOpen
  \bibfield  {author} {\bibinfo {author} {\bibfnamefont {H.~B.}\ \bibnamefont
  {Gray}},\ }\href@noop {} {\emph {\bibinfo {title} {Chemical Bonds: An
  Introduction to Atomic and Molecular Structure}}}\ (\bibinfo  {publisher}
  {{Univ. Science Books}},\ \bibinfo {address} {{Mill Valley, Calif}},\
  \bibinfo {year} {1994})\BibitemShut {NoStop}%
\bibitem [{\citenamefont {Veit}\ \emph {et~al.}(2021)\citenamefont {Veit},
  \citenamefont {Zuber}, \citenamefont {{Herrera-Sancho}}, \citenamefont
  {Anasuri}, \citenamefont {Schmid}, \citenamefont {Meinert}, \citenamefont
  {L{\"o}w},\ and\ \citenamefont {Pfau}}]{veitPulsedIonMicroscope2021}%
  \BibitemOpen
  \bibfield  {author} {\bibinfo {author} {\bibfnamefont {C.}~\bibnamefont
  {Veit}}, \bibinfo {author} {\bibfnamefont {N.}~\bibnamefont {Zuber}},
  \bibinfo {author} {\bibfnamefont {O.~A.}\ \bibnamefont {{Herrera-Sancho}}},
  \bibinfo {author} {\bibfnamefont {V.~S.~V.}\ \bibnamefont {Anasuri}},
  \bibinfo {author} {\bibfnamefont {T.}~\bibnamefont {Schmid}}, \bibinfo
  {author} {\bibfnamefont {F.}~\bibnamefont {Meinert}}, \bibinfo {author}
  {\bibfnamefont {R.}~\bibnamefont {L{\"o}w}},\ and\ \bibinfo {author}
  {\bibfnamefont {T.}~\bibnamefont {Pfau}},\ }\bibfield  {title} {\bibinfo
  {title} {Pulsed {{Ion Microscope}} to {{Probe Quantum Gases}}},\ }\href
  {https://doi.org/10.1103/PhysRevX.11.011036} {\bibfield  {journal} {\bibinfo
  {journal} {Physical Review X}\ }\textbf {\bibinfo {volume} {11}},\ \bibinfo
  {pages} {011036} (\bibinfo {year} {2021})}\BibitemShut {NoStop}%
\bibitem [{\citenamefont {K{\"o}hler}\ \emph {et~al.}(2006)\citenamefont
  {K{\"o}hler}, \citenamefont {G{\'o}ral},\ and\ \citenamefont
  {Julienne}}]{kohlerProductionColdMolecules2006}%
  \BibitemOpen
  \bibfield  {author} {\bibinfo {author} {\bibfnamefont {T.}~\bibnamefont
  {K{\"o}hler}}, \bibinfo {author} {\bibfnamefont {K.}~\bibnamefont
  {G{\'o}ral}},\ and\ \bibinfo {author} {\bibfnamefont {P.~S.}\ \bibnamefont
  {Julienne}},\ }\bibfield  {title} {\bibinfo {title} {Production of cold
  molecules via magnetically tunable {{Feshbach}} resonances},\ }\href
  {https://doi.org/10.1103/RevModPhys.78.1311} {\bibfield  {journal} {\bibinfo
  {journal} {Reviews of Modern Physics}\ }\textbf {\bibinfo {volume} {78}},\
  \bibinfo {pages} {1311} (\bibinfo {year} {2006})}\BibitemShut {NoStop}%
\bibitem [{\citenamefont {Greene}\ \emph {et~al.}(2017)\citenamefont {Greene},
  \citenamefont {Giannakeas},\ and\ \citenamefont
  {{P{\'e}rez-R{\'i}os}}}]{greeneUniversalFewbodyPhysics2017a}%
  \BibitemOpen
  \bibfield  {author} {\bibinfo {author} {\bibfnamefont {C.~H.}\ \bibnamefont
  {Greene}}, \bibinfo {author} {\bibfnamefont {P.}~\bibnamefont {Giannakeas}},\
  and\ \bibinfo {author} {\bibfnamefont {J.}~\bibnamefont
  {{P{\'e}rez-R{\'i}os}}},\ }\bibfield  {title} {\bibinfo {title} {Universal
  few-body physics and cluster formation},\ }\href
  {https://doi.org/10.1103/RevModPhys.89.035006} {\bibfield  {journal}
  {\bibinfo  {journal} {Reviews of Modern Physics}\ }\textbf {\bibinfo {volume}
  {89}},\ \bibinfo {pages} {035006} (\bibinfo {year} {2017})}\BibitemShut
  {NoStop}%
\bibitem [{\citenamefont {Nichols}\ \emph {et~al.}(2021)\citenamefont
  {Nichols}, \citenamefont {Liu}, \citenamefont {Zhu}, \citenamefont {Hu},
  \citenamefont {Liu},\ and\ \citenamefont
  {Ni}}]{nicholsDetectionLongLivedComplexes2021}%
  \BibitemOpen
  \bibfield  {author} {\bibinfo {author} {\bibfnamefont {M.~A.}\ \bibnamefont
  {Nichols}}, \bibinfo {author} {\bibfnamefont {Y.-X.}\ \bibnamefont {Liu}},
  \bibinfo {author} {\bibfnamefont {L.}~\bibnamefont {Zhu}}, \bibinfo {author}
  {\bibfnamefont {M.-G.}\ \bibnamefont {Hu}}, \bibinfo {author} {\bibfnamefont
  {Y.}~\bibnamefont {Liu}},\ and\ \bibinfo {author} {\bibfnamefont {K.-K.}\
  \bibnamefont {Ni}},\ }\bibfield  {title} {\bibinfo {title} {Detection of
  {{Long}}-{{Lived Complexes}} in {{Ultracold Atom}}-{{Molecule Collisions}}},\
  }\href@noop {} {\bibfield  {journal} {\bibinfo  {journal} {arXiv:2105.14960}\
  } (\bibinfo {year} {2021})},\ \Eprint {https://arxiv.org/abs/2105.14960}
  {arXiv:2105.14960} \BibitemShut {NoStop}%
\bibitem [{\citenamefont {Mayle}\ \emph {et~al.}(2013)\citenamefont {Mayle},
  \citenamefont {Qu{\'e}m{\'e}ner}, \citenamefont {Ruzic},\ and\ \citenamefont
  {Bohn}}]{mayleScatteringUltracoldMolecules2013}%
  \BibitemOpen
  \bibfield  {author} {\bibinfo {author} {\bibfnamefont {M.}~\bibnamefont
  {Mayle}}, \bibinfo {author} {\bibfnamefont {G.}~\bibnamefont
  {Qu{\'e}m{\'e}ner}}, \bibinfo {author} {\bibfnamefont {B.~P.}\ \bibnamefont
  {Ruzic}},\ and\ \bibinfo {author} {\bibfnamefont {J.~L.}\ \bibnamefont
  {Bohn}},\ }\bibfield  {title} {\bibinfo {title} {Scattering of ultracold
  molecules in the highly resonant regime},\ }\href
  {https://doi.org/10.1103/PhysRevA.87.012709} {\bibfield  {journal} {\bibinfo
  {journal} {Physical Review A}\ }\textbf {\bibinfo {volume} {87}},\ \bibinfo
  {pages} {012709} (\bibinfo {year} {2013})}\BibitemShut {NoStop}%
\bibitem [{\citenamefont {Shaffer}\ \emph {et~al.}(2018)\citenamefont
  {Shaffer}, \citenamefont {Rittenhouse},\ and\ \citenamefont
  {Sadeghpour}}]{shafferUltracoldRydbergMolecules2018}%
  \BibitemOpen
  \bibfield  {author} {\bibinfo {author} {\bibfnamefont {J.~P.}\ \bibnamefont
  {Shaffer}}, \bibinfo {author} {\bibfnamefont {S.~T.}\ \bibnamefont
  {Rittenhouse}},\ and\ \bibinfo {author} {\bibfnamefont {H.~R.}\ \bibnamefont
  {Sadeghpour}},\ }\bibfield  {title} {\bibinfo {title} {Ultracold {{Rydberg}}
  molecules},\ }\href {https://doi.org/10.1038/s41467-018-04135-6} {\bibfield
  {journal} {\bibinfo  {journal} {Nature Communications}\ }\textbf {\bibinfo
  {volume} {9}},\ \bibinfo {pages} {1965} (\bibinfo {year} {2018})}\BibitemShut
  {NoStop}%
\bibitem [{\citenamefont {Fey}\ \emph {et~al.}(2020)\citenamefont {Fey},
  \citenamefont {Hummel},\ and\ \citenamefont
  {Schmelcher}}]{feyUltralongrangeRydbergMolecules2020}%
  \BibitemOpen
  \bibfield  {author} {\bibinfo {author} {\bibfnamefont {C.}~\bibnamefont
  {Fey}}, \bibinfo {author} {\bibfnamefont {F.}~\bibnamefont {Hummel}},\ and\
  \bibinfo {author} {\bibfnamefont {P.}~\bibnamefont {Schmelcher}},\ }\bibfield
   {title} {\bibinfo {title} {Ultralong-range {{Rydberg}} molecules},\ }\href
  {https://doi.org/10.1080/00268976.2019.1679401} {\bibfield  {journal}
  {\bibinfo  {journal} {Molecular Physics}\ }\textbf {\bibinfo {volume}
  {118}},\ \bibinfo {pages} {e1679401} (\bibinfo {year} {2020})}\BibitemShut
  {NoStop}%
\bibitem [{\citenamefont {Lukin}\ \emph {et~al.}(2001)\citenamefont {Lukin},
  \citenamefont {Fleischhauer}, \citenamefont {Cote}, \citenamefont {Duan},
  \citenamefont {Jaksch}, \citenamefont {Cirac},\ and\ \citenamefont
  {Zoller}}]{lukinDipoleBlockadeQuantum2001a}%
  \BibitemOpen
  \bibfield  {author} {\bibinfo {author} {\bibfnamefont {M.~D.}\ \bibnamefont
  {Lukin}}, \bibinfo {author} {\bibfnamefont {M.}~\bibnamefont {Fleischhauer}},
  \bibinfo {author} {\bibfnamefont {R.}~\bibnamefont {Cote}}, \bibinfo {author}
  {\bibfnamefont {L.~M.}\ \bibnamefont {Duan}}, \bibinfo {author}
  {\bibfnamefont {D.}~\bibnamefont {Jaksch}}, \bibinfo {author} {\bibfnamefont
  {J.~I.}\ \bibnamefont {Cirac}},\ and\ \bibinfo {author} {\bibfnamefont
  {P.}~\bibnamefont {Zoller}},\ }\bibfield  {title} {\bibinfo {title} {Dipole
  {{Blockade}} and {{Quantum Information Processing}} in {{Mesoscopic Atomic
  Ensembles}}},\ }\href {https://doi.org/10.1103/PhysRevLett.87.037901}
  {\bibfield  {journal} {\bibinfo  {journal} {Physical Review Letters}\
  }\textbf {\bibinfo {volume} {87}},\ \bibinfo {pages} {037901} (\bibinfo
  {year} {2001})}\BibitemShut {NoStop}%
\bibitem [{\citenamefont {Urban}\ \emph {et~al.}(2009)\citenamefont {Urban},
  \citenamefont {Johnson}, \citenamefont {Henage}, \citenamefont {Isenhower},
  \citenamefont {Yavuz}, \citenamefont {Walker},\ and\ \citenamefont
  {Saffman}}]{urbanObservationRydbergBlockade2009a}%
  \BibitemOpen
  \bibfield  {author} {\bibinfo {author} {\bibfnamefont {E.}~\bibnamefont
  {Urban}}, \bibinfo {author} {\bibfnamefont {T.~A.}\ \bibnamefont {Johnson}},
  \bibinfo {author} {\bibfnamefont {T.}~\bibnamefont {Henage}}, \bibinfo
  {author} {\bibfnamefont {L.}~\bibnamefont {Isenhower}}, \bibinfo {author}
  {\bibfnamefont {D.~D.}\ \bibnamefont {Yavuz}}, \bibinfo {author}
  {\bibfnamefont {T.~G.}\ \bibnamefont {Walker}},\ and\ \bibinfo {author}
  {\bibfnamefont {M.}~\bibnamefont {Saffman}},\ }\bibfield  {title} {\bibinfo
  {title} {Observation of {{Rydberg}} blockade between two atoms},\ }\href
  {https://doi.org/10.1038/nphys1178} {\bibfield  {journal} {\bibinfo
  {journal} {Nature Physics}\ }\textbf {\bibinfo {volume} {5}},\ \bibinfo
  {pages} {110} (\bibinfo {year} {2009})}\BibitemShut {NoStop}%
\bibitem [{\citenamefont {Schwarzkopf}\ \emph {et~al.}(2013)\citenamefont
  {Schwarzkopf}, \citenamefont {Anderson}, \citenamefont {Thaicharoen},\ and\
  \citenamefont {Raithel}}]{schwarzkopfSpatialCorrelationsRydberg2013}%
  \BibitemOpen
  \bibfield  {author} {\bibinfo {author} {\bibfnamefont {A.}~\bibnamefont
  {Schwarzkopf}}, \bibinfo {author} {\bibfnamefont {D.~A.}\ \bibnamefont
  {Anderson}}, \bibinfo {author} {\bibfnamefont {N.}~\bibnamefont
  {Thaicharoen}},\ and\ \bibinfo {author} {\bibfnamefont {G.}~\bibnamefont
  {Raithel}},\ }\bibfield  {title} {\bibinfo {title} {Spatial correlations
  between {{Rydberg}} atoms in an optical dipole trap},\ }\href
  {https://doi.org/10.1103/PhysRevA.88.061406} {\bibfield  {journal} {\bibinfo
  {journal} {Physical Review A}\ }\textbf {\bibinfo {volume} {88}},\ \bibinfo
  {pages} {061406} (\bibinfo {year} {2013})}\BibitemShut {NoStop}%
\bibitem [{\citenamefont {Schwarzkopf}\ \emph {et~al.}(2011)\citenamefont
  {Schwarzkopf}, \citenamefont {Sapiro},\ and\ \citenamefont
  {Raithel}}]{schwarzkopfImagingSpatialCorrelations2011}%
  \BibitemOpen
  \bibfield  {author} {\bibinfo {author} {\bibfnamefont {A.}~\bibnamefont
  {Schwarzkopf}}, \bibinfo {author} {\bibfnamefont {R.~E.}\ \bibnamefont
  {Sapiro}},\ and\ \bibinfo {author} {\bibfnamefont {G.}~\bibnamefont
  {Raithel}},\ }\bibfield  {title} {\bibinfo {title} {Imaging {{Spatial
  Correlations}} of {{Rydberg Excitations}} in {{Cold Atom Clouds}}},\ }\href
  {https://doi.org/10.1103/PhysRevLett.107.103001} {\bibfield  {journal}
  {\bibinfo  {journal} {Physical Review Letters}\ }\textbf {\bibinfo {volume}
  {107}},\ \bibinfo {pages} {103001} (\bibinfo {year} {2011})}\BibitemShut
  {NoStop}%
\bibitem [{\citenamefont {Schau{\ss}}\ \emph {et~al.}(2012)\citenamefont
  {Schau{\ss}}, \citenamefont {Cheneau}, \citenamefont {Endres}, \citenamefont
  {Fukuhara}, \citenamefont {Hild}, \citenamefont {Omran}, \citenamefont
  {Pohl}, \citenamefont {Gross}, \citenamefont {Kuhr},\ and\ \citenamefont
  {Bloch}}]{schaussObservationSpatiallyOrdered2012}%
  \BibitemOpen
  \bibfield  {author} {\bibinfo {author} {\bibfnamefont {P.}~\bibnamefont
  {Schau{\ss}}}, \bibinfo {author} {\bibfnamefont {M.}~\bibnamefont {Cheneau}},
  \bibinfo {author} {\bibfnamefont {M.}~\bibnamefont {Endres}}, \bibinfo
  {author} {\bibfnamefont {T.}~\bibnamefont {Fukuhara}}, \bibinfo {author}
  {\bibfnamefont {S.}~\bibnamefont {Hild}}, \bibinfo {author} {\bibfnamefont
  {A.}~\bibnamefont {Omran}}, \bibinfo {author} {\bibfnamefont
  {T.}~\bibnamefont {Pohl}}, \bibinfo {author} {\bibfnamefont {C.}~\bibnamefont
  {Gross}}, \bibinfo {author} {\bibfnamefont {S.}~\bibnamefont {Kuhr}},\ and\
  \bibinfo {author} {\bibfnamefont {I.}~\bibnamefont {Bloch}},\ }\bibfield
  {title} {\bibinfo {title} {Observation of spatially ordered structures in a
  two-dimensional {{Rydberg}} gas},\ }\href
  {https://doi.org/10.1038/nature11596} {\bibfield  {journal} {\bibinfo
  {journal} {Nature}\ }\textbf {\bibinfo {volume} {491}},\ \bibinfo {pages}
  {87} (\bibinfo {year} {2012})}\BibitemShut {NoStop}%
\bibitem [{\citenamefont {Browaeys}\ and\ \citenamefont
  {Lahaye}(2020)}]{browaeysManybodyPhysicsIndividually2020}%
  \BibitemOpen
  \bibfield  {author} {\bibinfo {author} {\bibfnamefont {A.}~\bibnamefont
  {Browaeys}}\ and\ \bibinfo {author} {\bibfnamefont {T.}~\bibnamefont
  {Lahaye}},\ }\bibfield  {title} {\bibinfo {title} {Many-body physics with
  individually controlled {{Rydberg}} atoms},\ }\href
  {https://doi.org/10.1038/s41567-019-0733-z} {\bibfield  {journal} {\bibinfo
  {journal} {Nature Physics}\ }\textbf {\bibinfo {volume} {16}},\ \bibinfo
  {pages} {132} (\bibinfo {year} {2020})}\BibitemShut {NoStop}%
\bibitem [{\citenamefont {Hollerith}\ \emph {et~al.}(2019)\citenamefont
  {Hollerith}, \citenamefont {Zeiher}, \citenamefont {Rui}, \citenamefont
  {{Rubio-Abadal}}, \citenamefont {Walther}, \citenamefont {Pohl},
  \citenamefont {{Stamper-Kurn}}, \citenamefont {Bloch},\ and\ \citenamefont
  {Gross}}]{hollerithQuantumGasMicroscopy2019}%
  \BibitemOpen
  \bibfield  {author} {\bibinfo {author} {\bibfnamefont {S.}~\bibnamefont
  {Hollerith}}, \bibinfo {author} {\bibfnamefont {J.}~\bibnamefont {Zeiher}},
  \bibinfo {author} {\bibfnamefont {J.}~\bibnamefont {Rui}}, \bibinfo {author}
  {\bibfnamefont {A.}~\bibnamefont {{Rubio-Abadal}}}, \bibinfo {author}
  {\bibfnamefont {V.}~\bibnamefont {Walther}}, \bibinfo {author} {\bibfnamefont
  {T.}~\bibnamefont {Pohl}}, \bibinfo {author} {\bibfnamefont {D.~M.}\
  \bibnamefont {{Stamper-Kurn}}}, \bibinfo {author} {\bibfnamefont
  {I.}~\bibnamefont {Bloch}},\ and\ \bibinfo {author} {\bibfnamefont
  {C.}~\bibnamefont {Gross}},\ }\bibfield  {title} {\bibinfo {title} {Quantum
  gas microscopy of {{Rydberg}} macrodimers},\ }\href
  {https://doi.org/10.1126/science.aaw4150} {\bibfield  {journal} {\bibinfo
  {journal} {Science}\ }\textbf {\bibinfo {volume} {364}},\ \bibinfo {pages}
  {664} (\bibinfo {year} {2019})}\BibitemShut {NoStop}%
\bibitem [{\citenamefont {Hollerith}\ \emph {et~al.}(2021)\citenamefont
  {Hollerith}, \citenamefont {Rui}, \citenamefont {{Rubio-Abadal}},
  \citenamefont {Srakaew}, \citenamefont {Wei}, \citenamefont {Zeiher},
  \citenamefont {Gross},\ and\ \citenamefont
  {Bloch}}]{hollerithMicroscopicElectronicStructure2021b}%
  \BibitemOpen
  \bibfield  {author} {\bibinfo {author} {\bibfnamefont {S.}~\bibnamefont
  {Hollerith}}, \bibinfo {author} {\bibfnamefont {J.}~\bibnamefont {Rui}},
  \bibinfo {author} {\bibfnamefont {A.}~\bibnamefont {{Rubio-Abadal}}},
  \bibinfo {author} {\bibfnamefont {K.}~\bibnamefont {Srakaew}}, \bibinfo
  {author} {\bibfnamefont {D.}~\bibnamefont {Wei}}, \bibinfo {author}
  {\bibfnamefont {J.}~\bibnamefont {Zeiher}}, \bibinfo {author} {\bibfnamefont
  {C.}~\bibnamefont {Gross}},\ and\ \bibinfo {author} {\bibfnamefont
  {I.}~\bibnamefont {Bloch}},\ }\bibfield  {title} {\bibinfo {title}
  {Microscopic electronic structure tomography of {{Rydberg}} macrodimers},\
  }\href {https://doi.org/10.1103/PhysRevResearch.3.013252} {\bibfield
  {journal} {\bibinfo  {journal} {Physical Review Research}\ }\textbf {\bibinfo
  {volume} {3}},\ \bibinfo {pages} {013252} (\bibinfo {year}
  {2021})}\BibitemShut {NoStop}%
\bibitem [{\citenamefont {H{\"a}rter}\ and\ \citenamefont
  {Hecker~Denschlag}(2014)}]{harterColdAtomIon2014a}%
  \BibitemOpen
  \bibfield  {author} {\bibinfo {author} {\bibfnamefont {A.}~\bibnamefont
  {H{\"a}rter}}\ and\ \bibinfo {author} {\bibfnamefont {J.}~\bibnamefont
  {Hecker~Denschlag}},\ }\bibfield  {title} {\bibinfo {title} {Cold
  atom\textendash ion experiments in hybrid traps},\ }\href
  {https://doi.org/10.1080/00107514.2013.854618} {\bibfield  {journal}
  {\bibinfo  {journal} {Contemporary Physics}\ }\textbf {\bibinfo {volume}
  {55}},\ \bibinfo {pages} {33} (\bibinfo {year} {2014})}\BibitemShut {NoStop}%
\bibitem [{\citenamefont {Kleinbach}\ \emph {et~al.}(2018)\citenamefont
  {Kleinbach}, \citenamefont {Engel}, \citenamefont {Dieterle}, \citenamefont
  {L{\"o}w}, \citenamefont {Pfau},\ and\ \citenamefont
  {Meinert}}]{kleinbachIonicImpurityBoseEinstein2018}%
  \BibitemOpen
  \bibfield  {author} {\bibinfo {author} {\bibfnamefont {K.~S.}\ \bibnamefont
  {Kleinbach}}, \bibinfo {author} {\bibfnamefont {F.}~\bibnamefont {Engel}},
  \bibinfo {author} {\bibfnamefont {T.}~\bibnamefont {Dieterle}}, \bibinfo
  {author} {\bibfnamefont {R.}~\bibnamefont {L{\"o}w}}, \bibinfo {author}
  {\bibfnamefont {T.}~\bibnamefont {Pfau}},\ and\ \bibinfo {author}
  {\bibfnamefont {F.}~\bibnamefont {Meinert}},\ }\bibfield  {title} {\bibinfo
  {title} {Ionic {{Impurity}} in a {{Bose}}-{{Einstein Condensate}} at
  {{Submicrokelvin Temperatures}}},\ }\href
  {https://doi.org/10.1103/PhysRevLett.120.193401} {\bibfield  {journal}
  {\bibinfo  {journal} {Physical Review Letters}\ }\textbf {\bibinfo {volume}
  {120}},\ \bibinfo {pages} {193401} (\bibinfo {year} {2018})}\BibitemShut
  {NoStop}%
\bibitem [{\citenamefont {Tomza}\ \emph {et~al.}(2019)\citenamefont {Tomza},
  \citenamefont {Jachymski}, \citenamefont {Gerritsma}, \citenamefont
  {Negretti}, \citenamefont {Calarco}, \citenamefont {Idziaszek},\ and\
  \citenamefont {Julienne}}]{tomzaColdHybridIonatom2019b}%
  \BibitemOpen
  \bibfield  {author} {\bibinfo {author} {\bibfnamefont {M.}~\bibnamefont
  {Tomza}}, \bibinfo {author} {\bibfnamefont {K.}~\bibnamefont {Jachymski}},
  \bibinfo {author} {\bibfnamefont {R.}~\bibnamefont {Gerritsma}}, \bibinfo
  {author} {\bibfnamefont {A.}~\bibnamefont {Negretti}}, \bibinfo {author}
  {\bibfnamefont {T.}~\bibnamefont {Calarco}}, \bibinfo {author} {\bibfnamefont
  {Z.}~\bibnamefont {Idziaszek}},\ and\ \bibinfo {author} {\bibfnamefont
  {P.~S.}\ \bibnamefont {Julienne}},\ }\bibfield  {title} {\bibinfo {title}
  {Cold hybrid ion-atom systems},\ }\href
  {https://doi.org/10.1103/RevModPhys.91.035001} {\bibfield  {journal}
  {\bibinfo  {journal} {Reviews of Modern Physics}\ }\textbf {\bibinfo {volume}
  {91}},\ \bibinfo {pages} {035001} (\bibinfo {year} {2019})}\BibitemShut
  {NoStop}%
\bibitem [{\citenamefont {Ewald}\ \emph {et~al.}(2019)\citenamefont {Ewald},
  \citenamefont {Feldker}, \citenamefont {Hirzler}, \citenamefont {F{\"u}rst},\
  and\ \citenamefont {Gerritsma}}]{ewaldObservationInteractionsTrapped2019b}%
  \BibitemOpen
  \bibfield  {author} {\bibinfo {author} {\bibfnamefont {N.~V.}\ \bibnamefont
  {Ewald}}, \bibinfo {author} {\bibfnamefont {T.}~\bibnamefont {Feldker}},
  \bibinfo {author} {\bibfnamefont {H.}~\bibnamefont {Hirzler}}, \bibinfo
  {author} {\bibfnamefont {H.~A.}\ \bibnamefont {F{\"u}rst}},\ and\ \bibinfo
  {author} {\bibfnamefont {R.}~\bibnamefont {Gerritsma}},\ }\bibfield  {title}
  {\bibinfo {title} {Observation of {{Interactions}} between {{Trapped Ions}}
  and {{Ultracold Rydberg Atoms}}},\ }\href
  {https://doi.org/10.1103/PhysRevLett.122.253401} {\bibfield  {journal}
  {\bibinfo  {journal} {Physical Review Letters}\ }\textbf {\bibinfo {volume}
  {122}},\ \bibinfo {pages} {253401} (\bibinfo {year} {2019})}\BibitemShut
  {NoStop}%
\bibitem [{\citenamefont {Dieterle}\ \emph {et~al.}(2020)\citenamefont
  {Dieterle}, \citenamefont {Berngruber}, \citenamefont {H{\"o}lzl},
  \citenamefont {L{\"o}w}, \citenamefont {Jachymski}, \citenamefont {Pfau},\
  and\ \citenamefont {Meinert}}]{dieterleInelasticCollisionDynamics2020}%
  \BibitemOpen
  \bibfield  {author} {\bibinfo {author} {\bibfnamefont {T.}~\bibnamefont
  {Dieterle}}, \bibinfo {author} {\bibfnamefont {M.}~\bibnamefont
  {Berngruber}}, \bibinfo {author} {\bibfnamefont {C.}~\bibnamefont
  {H{\"o}lzl}}, \bibinfo {author} {\bibfnamefont {R.}~\bibnamefont {L{\"o}w}},
  \bibinfo {author} {\bibfnamefont {K.}~\bibnamefont {Jachymski}}, \bibinfo
  {author} {\bibfnamefont {T.}~\bibnamefont {Pfau}},\ and\ \bibinfo {author}
  {\bibfnamefont {F.}~\bibnamefont {Meinert}},\ }\bibfield  {title} {\bibinfo
  {title} {Inelastic collision dynamics of a single cold ion immersed in a
  {{Bose}}-{{Einstein}} condensate},\ }\href
  {https://doi.org/10.1103/PhysRevA.102.041301} {\bibfield  {journal} {\bibinfo
   {journal} {Physical Review A}\ }\textbf {\bibinfo {volume} {102}},\ \bibinfo
  {pages} {041301} (\bibinfo {year} {2020})}\BibitemShut {NoStop}%
\bibitem [{\citenamefont {Schmidt}\ \emph {et~al.}(2020)\citenamefont
  {Schmidt}, \citenamefont {Weckesser}, \citenamefont {Thielemann},
  \citenamefont {Schaetz},\ and\ \citenamefont
  {Karpa}}]{schmidtOpticalTrapsSympathetic2020}%
  \BibitemOpen
  \bibfield  {author} {\bibinfo {author} {\bibfnamefont {J.}~\bibnamefont
  {Schmidt}}, \bibinfo {author} {\bibfnamefont {P.}~\bibnamefont {Weckesser}},
  \bibinfo {author} {\bibfnamefont {F.}~\bibnamefont {Thielemann}}, \bibinfo
  {author} {\bibfnamefont {T.}~\bibnamefont {Schaetz}},\ and\ \bibinfo {author}
  {\bibfnamefont {L.}~\bibnamefont {Karpa}},\ }\bibfield  {title} {\bibinfo
  {title} {Optical {{Traps}} for {{Sympathetic Cooling}} of {{Ions}} with
  {{Ultracold Neutral Atoms}}},\ }\href
  {https://doi.org/10.1103/PhysRevLett.124.053402} {\bibfield  {journal}
  {\bibinfo  {journal} {Physical Review Letters}\ }\textbf {\bibinfo {volume}
  {124}},\ \bibinfo {pages} {053402} (\bibinfo {year} {2020})}\BibitemShut
  {NoStop}%
\bibitem [{\citenamefont {Dieterle}\ \emph {et~al.}(2021)\citenamefont
  {Dieterle}, \citenamefont {Berngruber}, \citenamefont {H{\"o}lzl},
  \citenamefont {L{\"o}w}, \citenamefont {Jachymski}, \citenamefont {Pfau},\
  and\ \citenamefont {Meinert}}]{dieterleTransportSingleCold2021}%
  \BibitemOpen
  \bibfield  {author} {\bibinfo {author} {\bibfnamefont {T.}~\bibnamefont
  {Dieterle}}, \bibinfo {author} {\bibfnamefont {M.}~\bibnamefont
  {Berngruber}}, \bibinfo {author} {\bibfnamefont {C.}~\bibnamefont
  {H{\"o}lzl}}, \bibinfo {author} {\bibfnamefont {R.}~\bibnamefont {L{\"o}w}},
  \bibinfo {author} {\bibfnamefont {K.}~\bibnamefont {Jachymski}}, \bibinfo
  {author} {\bibfnamefont {T.}~\bibnamefont {Pfau}},\ and\ \bibinfo {author}
  {\bibfnamefont {F.}~\bibnamefont {Meinert}},\ }\bibfield  {title} {\bibinfo
  {title} {Transport of a {{Single Cold Ion Immersed}} in a {{Bose}}-{{Einstein
  Condensate}}},\ }\href {https://doi.org/10.1103/PhysRevLett.126.033401}
  {\bibfield  {journal} {\bibinfo  {journal} {Physical Review Letters}\
  }\textbf {\bibinfo {volume} {126}},\ \bibinfo {pages} {033401} (\bibinfo
  {year} {2021})}\BibitemShut {NoStop}%
\bibitem [{\citenamefont {Weckesser}\ \emph {et~al.}(2021)\citenamefont
  {Weckesser}, \citenamefont {Thielemann}, \citenamefont {Wiater},
  \citenamefont {Wojciechowska}, \citenamefont {Karpa}, \citenamefont
  {Jachymski}, \citenamefont {Tomza}, \citenamefont {Walker},\ and\
  \citenamefont {Schaetz}}]{weckesserObservationFeshbachResonances2021}%
  \BibitemOpen
  \bibfield  {author} {\bibinfo {author} {\bibfnamefont {P.}~\bibnamefont
  {Weckesser}}, \bibinfo {author} {\bibfnamefont {F.}~\bibnamefont
  {Thielemann}}, \bibinfo {author} {\bibfnamefont {D.}~\bibnamefont {Wiater}},
  \bibinfo {author} {\bibfnamefont {A.}~\bibnamefont {Wojciechowska}}, \bibinfo
  {author} {\bibfnamefont {L.}~\bibnamefont {Karpa}}, \bibinfo {author}
  {\bibfnamefont {K.}~\bibnamefont {Jachymski}}, \bibinfo {author}
  {\bibfnamefont {M.}~\bibnamefont {Tomza}}, \bibinfo {author} {\bibfnamefont
  {T.}~\bibnamefont {Walker}},\ and\ \bibinfo {author} {\bibfnamefont
  {T.}~\bibnamefont {Schaetz}},\ }\bibfield  {title} {\bibinfo {title}
  {Observation of {{Feshbach}} resonances between a single ion and ultracold
  atoms},\ }\href@noop {} {\bibfield  {journal} {\bibinfo  {journal}
  {arXiv:2105.09382}\ } (\bibinfo {year} {2021})},\ \Eprint
  {https://arxiv.org/abs/2105.09382} {arXiv:2105.09382} \BibitemShut {NoStop}%
\bibitem [{\citenamefont {Duspayev}\ \emph {et~al.}(2021)\citenamefont
  {Duspayev}, \citenamefont {Han}, \citenamefont {Viray}, \citenamefont {Ma},
  \citenamefont {Zhao},\ and\ \citenamefont
  {Raithel}}]{duspayevLongrangeRydbergatomIon2021}%
  \BibitemOpen
  \bibfield  {author} {\bibinfo {author} {\bibfnamefont {A.}~\bibnamefont
  {Duspayev}}, \bibinfo {author} {\bibfnamefont {X.}~\bibnamefont {Han}},
  \bibinfo {author} {\bibfnamefont {M.~A.}\ \bibnamefont {Viray}}, \bibinfo
  {author} {\bibfnamefont {L.}~\bibnamefont {Ma}}, \bibinfo {author}
  {\bibfnamefont {J.}~\bibnamefont {Zhao}},\ and\ \bibinfo {author}
  {\bibfnamefont {G.}~\bibnamefont {Raithel}},\ }\bibfield  {title} {\bibinfo
  {title} {Long-range {{Rydberg}}-atom\textendash ion molecules of {{Rb}} and
  {{Cs}}},\ }\href {https://doi.org/10.1103/PhysRevResearch.3.023114}
  {\bibfield  {journal} {\bibinfo  {journal} {Physical Review Research}\
  }\textbf {\bibinfo {volume} {3}},\ \bibinfo {pages} {023114} (\bibinfo {year}
  {2021})}\BibitemShut {NoStop}%
\bibitem [{\citenamefont {Dei{\ss}}\ \emph {et~al.}(2021)\citenamefont
  {Dei{\ss}}, \citenamefont {Haze},\ and\ \citenamefont
  {Hecker~Denschlag}}]{deissLongRangeAtomIon2021}%
  \BibitemOpen
  \bibfield  {author} {\bibinfo {author} {\bibfnamefont {M.}~\bibnamefont
  {Dei{\ss}}}, \bibinfo {author} {\bibfnamefont {S.}~\bibnamefont {Haze}},\
  and\ \bibinfo {author} {\bibfnamefont {J.}~\bibnamefont {Hecker~Denschlag}},\
  }\bibfield  {title} {\bibinfo {title} {Long-{{Range Atom}}\textendash{{Ion
  Rydberg Molecule}}: A {{Novel Molecular Binding Mechanism}}},\ }\href
  {https://doi.org/10.3390/atoms9020034} {\bibfield  {journal} {\bibinfo
  {journal} {Atoms}\ }\textbf {\bibinfo {volume} {9}},\ \bibinfo {pages} {34}
  (\bibinfo {year} {2021})}\BibitemShut {NoStop}%
\bibitem [{\citenamefont {Duspayev}\ and\ \citenamefont
  {Raithel}(2021)}]{duspayevNonadiabaticDecayRydbergatomion2021}%
  \BibitemOpen
  \bibfield  {author} {\bibinfo {author} {\bibfnamefont {A.}~\bibnamefont
  {Duspayev}}\ and\ \bibinfo {author} {\bibfnamefont {G.}~\bibnamefont
  {Raithel}},\ }\bibfield  {title} {\bibinfo {title} {Non-adiabatic decay of
  {{Rydberg}}-atom-ion molecules},\ }\href@noop {} {\bibfield  {journal}
  {\bibinfo  {journal} {arXiv:2108.10475}\ } (\bibinfo {year} {2021})},\
  \Eprint {https://arxiv.org/abs/2108.10475} {arXiv:2108.10475} \BibitemShut
  {NoStop}%
\bibitem [{\citenamefont {Bonin}\ and\ \citenamefont
  {McIlrath}(1984)}]{boninTwophotonElectricdipoleSelection1984}%
  \BibitemOpen
  \bibfield  {author} {\bibinfo {author} {\bibfnamefont {K.~D.}\ \bibnamefont
  {Bonin}}\ and\ \bibinfo {author} {\bibfnamefont {T.~J.}\ \bibnamefont
  {McIlrath}},\ }\bibfield  {title} {\bibinfo {title} {Two-photon
  electric-dipole selection rules},\ }\href
  {https://doi.org/10.1364/JOSAB.1.000052} {\bibfield  {journal} {\bibinfo
  {journal} {JOSA B}\ }\textbf {\bibinfo {volume} {1}},\ \bibinfo {pages} {52}
  (\bibinfo {year} {1984})}\BibitemShut {NoStop}%
\bibitem [{SIM()}]{SIMION32}%
  \BibitemOpen
  \href@noop {} {\bibinfo {title} {{{SIMION}}, 8.1.1.32}},\ \bibinfo
  {howpublished} {Scientific Instrument Services}\BibitemShut {NoStop}%
\bibitem [{\citenamefont
  {Jackson}(1999)}]{jacksonClassicalElectrodynamics1999}%
  \BibitemOpen
  \bibfield  {author} {\bibinfo {author} {\bibfnamefont {J.~D.}\ \bibnamefont
  {Jackson}},\ }\href@noop {} {\emph {\bibinfo {title} {Classical
  Electrodynamics}}},\ \bibinfo {edition} {3rd}\ ed.\ (\bibinfo  {publisher}
  {{Wiley}},\ \bibinfo {address} {{New York}},\ \bibinfo {year}
  {1999})\BibitemShut {NoStop}%
\bibitem [{\citenamefont {Weber}\ \emph {et~al.}(2017)\citenamefont {Weber},
  \citenamefont {Tresp}, \citenamefont {Menke}, \citenamefont {Urvoy},
  \citenamefont {Firstenberg}, \citenamefont {B{\"u}chler},\ and\ \citenamefont
  {Hofferberth}}]{weberCalculationRydbergInteraction2017}%
  \BibitemOpen
  \bibfield  {author} {\bibinfo {author} {\bibfnamefont {S.}~\bibnamefont
  {Weber}}, \bibinfo {author} {\bibfnamefont {C.}~\bibnamefont {Tresp}},
  \bibinfo {author} {\bibfnamefont {H.}~\bibnamefont {Menke}}, \bibinfo
  {author} {\bibfnamefont {A.}~\bibnamefont {Urvoy}}, \bibinfo {author}
  {\bibfnamefont {O.}~\bibnamefont {Firstenberg}}, \bibinfo {author}
  {\bibfnamefont {H.~P.}\ \bibnamefont {B{\"u}chler}},\ and\ \bibinfo {author}
  {\bibfnamefont {S.}~\bibnamefont {Hofferberth}},\ }\bibfield  {title}
  {\bibinfo {title} {Calculation of {{Rydberg}} interaction potentials},\
  }\href {https://doi.org/10.1088/1361-6455/aa743a} {\bibfield  {journal}
  {\bibinfo  {journal} {Journal of Physics B: Atomic, Molecular and Optical
  Physics}\ }\textbf {\bibinfo {volume} {50}},\ \bibinfo {pages} {133001}
  (\bibinfo {year} {2017})}\BibitemShut {NoStop}%
\bibitem [{\citenamefont {Shore}(1990)}]{shoreTheoryCoherentAtomic1990}%
  \BibitemOpen
  \bibfield  {author} {\bibinfo {author} {\bibfnamefont {B.~W.}\ \bibnamefont
  {Shore}},\ }\href@noop {} {\emph {\bibinfo {title} {The Theory of Coherent
  Atomic Excitation}}}\ (\bibinfo  {publisher} {{Wiley}},\ \bibinfo {address}
  {{New York}},\ \bibinfo {year} {1990})\BibitemShut {NoStop}%
\end{thebibliography}
%
%
\appendix
\setcounter{figure}{0}
\renewcommand{\figurename}{Supplement FIG.}
\renewcommand\thefigure{S.\arabic{figure}}  
\renewcommand{\theHfigure}{Methods.\thefigure}
\section{Methods}
\subsection{Cloud preparation}
In every experimental cycle about $\SI{3e8}{Rb}$ atoms were trapped and cooled in a magneto optical trap (MOT) after slowing them down with a Zeeman slower from an effusive oven. To transport the atoms from the MOT region to the science chamber below the ion microscope, an optical transport was used that consisted of a \SI{1064}{nm} laser beam. A lens mounted on an air-bearing translation stage shifted the focus with a $1/e^2$ waist of \SI{38}{\micro m} along the $x$-axis. In the science chamber, the cold atoms were transferred to a crossed dipole trap by simultaneously reducing the transport beam power and increasing the power of a second dipole trap beam oriented along the $y$-axis within \SI{5.9}{s}. The waist of the second dipole trap was about \SI{40}{\micro m}. We worked at cloud densities between $\SI{1e12}{cm^{-3}}$ and $\SI{5e12}{cm^{-3}}$. For all the experiments carried out in this paper, the ground state atoms were pumped to the $F=2$ state, but were not spin-polarised. One experimental cycle took \SI{21}{s}.

Stray electric fields were compensated by adjusting the voltages applied to six electrodes. This was accomplished by observing the motion of single free ions in the object plane for a wait time of \SI{70}{\micro s} and minimizing the displacement of the ion distribution~\cite{veitPulsedIonMicroscope2021}. Along the optical axis the change in the time of flight was minimized. After the compensation procedure, the ion movement implies a residual electric field of a few $\SI{100}{\micro V/cm}$ at the position of the ions.

\section{Photoionization and photoassociation}
To photoassociate the molecule, we first created a cold ion via a two-photon ionization scheme. The first ionization laser had a wavelength of \SI{420}{nm} and pointed along the $y$-axis. The laser beam had a $1/\mathrm{e}^2$ waist of $\omega_{420}=\SI{7}{\micro m}$. The second laser was orientated along the $z$-axis with a waist of $\omega_{1010}=\SI{3.2}{\micro m}$ and a wavelength of \SI{1010}{nm}. The intermediate detuning $\Delta_{i_1}$ from the $\ket{6P_{3/2},F=3}$ state was about $\SI{80}{MHz}$ to the blue (Fig.~\ref{fig:concept}c). The ionization light pulses in the experiment were \SI{1}{\micro s} long, with both lasers switched simultaneously. Within $6{,}080$ ionization pulses about 600 to $1{,}200$ ions were detected. 

The Rydberg atoms were excited with a \SI{780}{nm} laser pointing along the $z$-axis with a waist much larger than our atomic cloud and a \SI{480}{nm} laser focused to a waist of $\omega_{480}=\SI{6}{\micro m}$ and pointing along the $y$-axis. This ensured that the excitation volume is larger than the molecular bond length, but still allows the excitation of up to 200 detected molecules within the $6{,}080$ pulses. The \SI{780}{nm} laser was detuned to the blue from the intermediate $\ket{5P_{3/2},F=3}$ state by \SI{250}{MHz}. In all the measurements, the duration of the photoassociation pulses was \SI{1.5}{\micro s} long, such that the excitation bandwidth of the laser pulse did not broaden the measured spectral width of the vibrational levels.

From the ratio of detected molecules and detected Rydberg cores we determine a lower bound of the detector efficiency of $\SI{47}{\%} \pm \SI{3}{\%}$. Taking the finite detection efficiency into account, the probability to create two ions in one measurement cycle simultaneously is below \SI{10}{\%}. If such a double ion event is detected it is filtered out during post processing. So we do  expect that the influence of double ion counts is negligible in the presented measurements.

\subsection{Separation pulse}
The imaging of the molecule consisted of a two step sequence of carefully adjusted electric field pulses. The first pulse separated the ion from the Rydberg atom along the optical axis before the imaging pulse field ionized the Rydberg atom and extracted both particles into the ion microscope.

We performed a characterization measurement with the vibrational ground state bound in the potential corresponding to the $\ket{69P_{1/2}}$ state at large internuclear distance and additional classical two particle trajectory simulations to estimate the influence of the separation pulse onto our in situ image. The characterization has been carried out with different separation pulse heights ranging from \SI{4.3}{V/cm} to \SI{8.6}{V/cm}. This test aims to ensure that the final pulse height, and the rise time of the pulse associated with, impose only small distortions onto the image. This is confirmed by the measurements where only minor changes in the shape of the resulting image were observed for a separation pulse duration of \SI{700}{ns}. Thus, the pulse height was fixed to \SI{6.9}{V/cm} for the in situ measurements with an initial rising flank of the pulse of \SI{2.6}{V/cm} within \SI{20}{ns}. For the spectroscopic measurements at the principle quantum number of $n=54$, the separation pulse height was increased to \SI{11.4}{V/cm} accompanied by a reduction of the pulse length to \SI{560}{ns}.

Additionally, we performed classical trajectory simulations to estimate the effect of the separation pulse onto the measured ion and Rydberg core positions. The simulations assumed an interaction potential, which corresponds to the molecular potential for zero external electric field. To approximate the transition from a bound molecule to a free evolution of each constituent, the interaction potential in the simulation was switched off when the raising external electric field pulse reached the same value as the field of the ion at the binding length. For the vibrational ground state in the potential below the $\ket{54P_{1/2}}$ asymptote the simulations show only minor distortions of the in situ image for rise times of the separation pulse of \SI{3}{V/cm} within \SI{20}{ns}. However, for slower pulses the radial distribution in the image becomes distorted. For higher principle quantum numbers of the Rydberg state, the requirements to the rise time are reduced. Therefore the main quantum number of the Rydberg state is increased to $n=69$ for the in situ images.

The difference in the time of flight of the ion and the Rydberg atom allows the estimation of the distance between the two particles after the separation pulse. We estimate the distance to be on the order of \SI{300}{\micro m} to \SI{400}{\micro m} along the optical axis. These conditions require a large depth of field of the ion microscope to measure the Rydberg atom and ion positions accurately. A depth of field of at least \SI{70}{\micro m} with a resolution of \SI{200}{nm} was already demonstrated in Ref.~\cite{veitPulsedIonMicroscope2021} for our ion microscope. Charged particle trajectory simulations~\cite{SIMION32} of the ion microscope show a high resolution over a range of up to $\pm\SI{1}{mm}$ along the optical axis and a deviation of the magnification below $5\%$ for positions separated by \SI{400}{\micro m} can be expected. The simulations were verified by imaging a small volume of ions with and without the separation pulse. The waists of the imaged distribution of $\omega_x = \SI{3.7}{\micro m}\pm\SI{0.1}{\micro m}$ and $\omega_z = \SI{4.7}{\micro m}\pm\SI{0.1}{\micro m}$ did not change within the errors bars. Hence, we do not expect a significant influence of the large separation between the ion and Rydberg atom along the optical axis onto the measured ion and Rydberg atom positions.

\subsection{Vibrational ground state below the $69P_{1/2}$ asymptote}
\begin{figure}[t]
\centering
\includegraphics[width=\columnwidth]{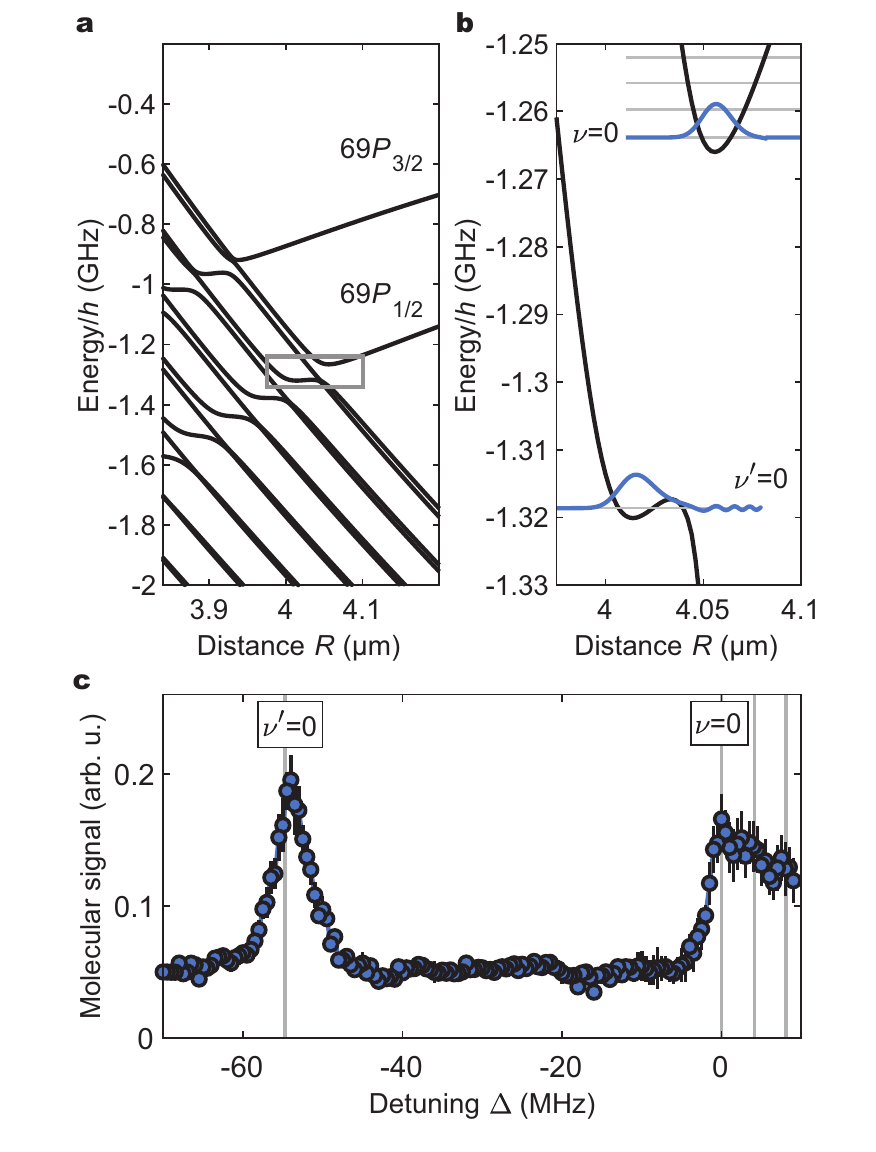}
\caption{\textbf{Potential energy curves with bound vibrational states and measured spectrum.} \textbf{a}, Potential energy curves corresponding to the $69P$-state with $|m_J|=1/2$. \textbf{b}, Magnified view of the region marked with a grey box in \textbf{a} with the two potential curves containing the vibrational bound states. Grey lines indicate the energy of vibrational states with vibrational wave functions plotted in blue. \textbf{c}, Normalised molecular signal as a function of the Rydberg laser detuning $\Delta$. Error bars denote standard error of the mean.}
\label{fig:spectrum69P}%
\end{figure}

\begin{figure}[t]
\centering
\includegraphics[width=\columnwidth]{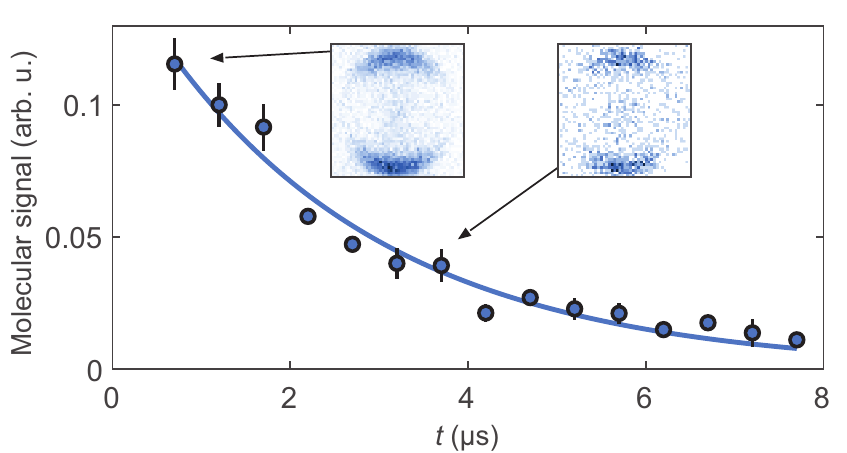}
\caption{\textbf{Lifetime measurement with in situ images.} Molecular signal strength measured for different wait times $t$ between the molecule creation and imaging. The exponential fit (blue line) results in a 1/e lifetime of $\SI{2.6}{\micro s}\pm\SI{0.2}{\micro s}$ for the lowest vibrational state in the molecular potential corresponding to the $\ket{69P_{1/2}}$-state. The insets show two in situ images for \SI{0.7}{\micro s} and \SI{3.5}{\micro s} wait time. Error bars denote one standard deviation.}
\label{fig:lifetimeInsituImage}%
\end{figure}

To identify the vibrational ground state in the potential corresponding to the $\ket{69P_{1/2}}$ asymptote, we recorded the spectrum shown in Fig.~\ref{fig:spectrum69P}. As in the measurement for Fig.~\ref{fig:spectrum}, the Rydberg lasers polarisations were orientated parallel to each other. The theoretically calculated bound states in Fig.~\ref{fig:spectrum69P}b predict vibrational states with a maximum splitting of about \SI{4}{MHz}. Consequently, the vibrational states are resolved with a lower contrast compared to Fig.~\ref{fig:spectrum}. To suppress the excitation of higher vibrational levels, we set our excitation laser about \SI{1}{MHz} red detuned to the lowest vibrational level $\nu=0$. We image the molecule after an additional wait time of $\SI{3.5}{\micro s}$ following its photoassociation (see Fig.~\ref{fig:lifetimeInsituImage}). The wait time exceeds the extracted lifetime of $\SI{2.6}{\micro s}\pm\SI{0.2}{\micro s}$. Even with this additional wait time radially confined Rydberg cores with respect to the ion are visible, confirming that we excite to a bound molecular state.

\subsection{Potential energy curves calculations}

The theory of the Rydberg-atom-ion molecules is motivated by the references~\cite{duspayevLongrangeRydbergatomIon2021,deissLongRangeAtomIon2021}. We refer to these  for the full details, but give a short summary of the calculations here. The internuclear distance $\vect{R}$ between the ion and the Rydberg atom is assumed to be much larger than the size of the Rydberg atom. For these large distances and the studied Rydberg states, internuclear tunnelling of the Rydberg electron is suppressed~\cite{deissLongRangeAtomIon2021}. The Hamiltonian representing the system $H = H_0 + V_I$ consists of a part $H_0$ describing the unperturbed Rydberg atom and a term $V_I$ describing the interaction between a point like ion and the Rydberg atom. The latter is expressed in a multipole expansion~\cite{jacksonClassicalElectrodynamics1999, weberCalculationRydbergInteraction2017}
\begin{equation*}
V_I = -\frac{e^2}{4\pi\epsilon_0} \sum_{\kappa=1}^\infty \sqrt{ \frac{4\pi}{2\kappa+1}} \frac{r_e^\kappa}{R^{\kappa+1}} Y_{\kappa,0}(\theta_e,\phi_e) \, .
\end{equation*}
Here $r_e$ is the position of the Rydberg electron with respect to the Rydberg core, $Y_{\kappa,0}(\theta_e,\phi_e)$ are the spherical harmonics, which depend on the angular position of the Rydberg electron, and $\kappa$ is the order of the multipole interaction. The vacuum permittivity is $\epsilon_0$ and the elementary charge is $e$. The monopole interaction $\kappa=0$ vanishes for the neutral Rydberg atom. The maximum multipole order taken into account is $\kappa=6$ as it has been shown in Ref.~~\cite{duspayevLongrangeRydbergatomIon2021} that higher orders have negligible effect on the calculated potentials. The potential energy curves are calculated in Born-Oppenheimer approximation by using the unperturbed Rydberg states as basis states to numerically construct and diagonalize the Hamiltonian for fixed internuclear separations $R$. We use the pairinteraction programme~\cite{weberCalculationRydbergInteraction2017} to calculate the potential energy curves, which was modified to include the multipolar interaction with a point charge. The basis states are truncated by introducing a lower and upper limit in the principle quantum number $n$ of $n_{\mathrm{min}} = n-6$ and $n_{\mathrm{max}} = n+6$. 

\subsection{Two-photon excitation probability}
\begin{figure}[tb]
\centering
\includegraphics[width=\columnwidth]{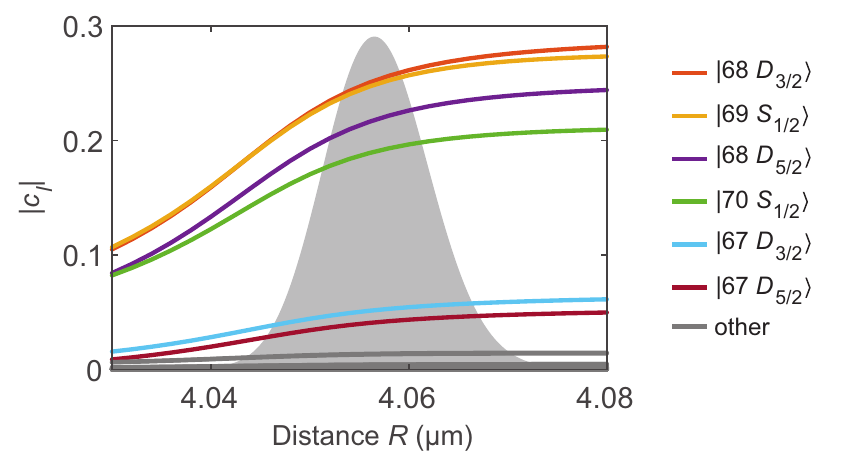}
\caption{\textbf{Admixture of Ryberg states.} Contributions $|c_l|$ of states mixed into the electronic molecular state below the $\ket{69P_{1/2}}$ asymptote. Here only those states are shown which can be excited with the two-photon scheme from the ground state $\ket{g}$. In calculation of the two-photon excitation probability the electronic molecular state is truncated after the first six states with the largest contribution (coloured). The nuclear radial probability density (light grey) indicates the extend of the lowest vibrational state $\nu=0$.}
\label{fig:prefactors69P}%
\end{figure}

A single rotational state cannot be resolved spectroscopically during the lifetime of the molecule as the rotational constant for the $54P$-state is on the order of \SI{26}{Hz} and for the $69P$-state around \SI{7}{Hz}. In combination with the low temperature of our atoms and ions, we assume the orientation of the molecular axis to be fixed during the observation times used in this paper. For a homogeneous background gas around the ion, the probability to excite the Rydberg atom in a certain direction with respect to the ion is given by the two-photon excitation probability per unit time~\cite{boninTwophotonElectricdipoleSelection1984,shoreTheoryCoherentAtomic1990} to go from the ground state $\ket{g}$ to the excited state $\ket{e}$
\begin{equation*}
P_{g\rightarrow e}\propto \left| \sum_{i_2} \frac{\braket{e|\vect{\varepsilon}_{480} \cdot \hat{\vect{d}} |i_2} \braket{i_2|\vect{\varepsilon}_{780} \cdot \hat{\vect{d}} |g}}{\Delta_{i_2}} \right|^2 \, .
\end{equation*}
The natural linewidth $\Gamma$ of the intermediate states is much smaller than the detuning $\Delta_{i_2}$ of the excitation lasers to the intermediate states $\ket{i_2}$ and is therefore neglected in the calculations. Here $i_2$ denotes the different hyperfine states of the intermediate $\ket{5P_{3/2},F,m_F}$ states. The laser polarisations $\vect{\varepsilon}_\lambda$ for the two Rydberg lasers with wavelength $\lambda$ are defined in the molecular reference frame with the quantization axis along the molecular axis and $\hat{\vect{d}}$ is the dipole operator. The orientation of the molecule with respect to the lasers polarisation causes an angular dependence of the excitation probability. The excited state $\ket{e} = \sum_{l} c_l \ket{l}$ is a linear combination of the different unperturbed Rydberg states. Here $l$ denotes all the quantum numbers of the fine structure Rydberg states and coefficients $c_l$ denote the admixture to the excited electronic state. The two-photon Rydberg excitation lasers only couple to the $S$- and $D$-state components of the excited state. From all the possible $S$- and $D$-components in the excited state only six states contribute significantly (Fig.~\ref{fig:prefactors69P}) and are considered in the calculations. In addition, the $R$ dependence of the coefficients $c_l$ over the extend of the vibrational ground state is neglected in the calculations. The ground state is an incoherent mixture of all the $\ket{5S_{1/2},F=2,m_F}$ hyperfine states and the excitation probabilities can be summed up over all $m_F$ states of the ground state. The probabilities to excite one of the two excited states with $m_J=+1/2$ or $m_J=-1/2$ are calculated independently and are summed up resulting in
\begin{equation*}
P_{tot}\propto \sum_{m_F,m_J} \left| \sum_{i_2} \frac{\braket{e,m_J|\vect{\varepsilon}_{480} \cdot \hat{\vect{d}} |i_2} \braket{i_2|\vect{\varepsilon}_{780} \cdot \hat{\vect{d}} |g,m_F}}{\Delta_{i_2}} \right|^2 \, .
\end{equation*}

\subsection{Data processing}
The detector attributes two spatial and one temporal coordinate to each detected particle. The temporal data reflects the time of flight of the particle through our ion microscope and allows to distinguish between Rydberg cores and ions due to their different time of flight. For data analysis, we define a separation time that determines two detection windows. Events with a shorter time stamp are attributed as Rydberg atoms, while detected particles with a larger time of flight are assigned as ions. This separation time needs to be adjusted depending on the magnification of the ion microscope and the separation pulse parameters. In post processing the molecular events were identified by selecting only events, where exactly one particle was detected in each of the two detection windows. In a similar manner, single ion events or single Rydberg events can be identified. The spatial coordinates given by the detector were converted to coordinates in the object space by the known magnifications for calibrated voltage settings of the ion microscope~\cite{veitPulsedIonMicroscope2021}.

\end{document}